\documentclass[prl,amsmath,amssymb,twocolumn,superscriptaddress]{revtex4}
\usepackage{graphicx}

\newcommand{\bk}{{\bf k}}

\newcommand{\bp}{{\bf p}}
\newcommand{\bq}{{\bf q}}

\newcommand{\bE}{{\bf E}}
\newcommand{\bS}{{\bf S}}
\newcommand{\bM}{{\bf M}}

\newcommand{\eps}{\epsilon}

\begin{document}
\title{Piezo-Magneto-Electric Effects in p-Doped Semiconductors}
\author{B.Andrei Bernevig and Oskar Vafek}
\affiliation{Department of Physics, Stanford University, CA 94305}
\begin{abstract}
We predict the appearance of a uniform magnetization in strained three
dimensional p-doped semiconductors with inversion symmetry
breaking subject to an external electric field. We compute
the magnetization response to the electric field as a function of
the direction and magnitude of the applied strain. This
effect could be used to manipulate the collective magnetic moment of hole
mediated ferromagnetism of magnetically doped semiconductors.
\end{abstract} \maketitle

Antiferromagnetic dielectrics with inversion asymmetry exhibit the
magnetoelectric (ME) effect, a phenomenon in which a static electric
field induces a uniform magnetization \cite{landau,schmidt75}.
Moreover, as first pointed out by Levitov {\it et.al.}
\cite{levitov}, a kinematic magnetoelectric (kME) effect can also
occur in ostensibly nonmagnetic {\em conductors}, with
spin-orbit coupling, which lack a center
of inversion symmetry. Unlike
in dielectrics, in the case of conductors, the electric field
induced magnetization density, $M_{i}=\alpha_{ij} E_{j}$, is
necessarily accompanied by dissipation. Since $\bM$ is odd under
time reversal (T) and $\bE$ is even, $\alpha_{ij}$ must be proportional
to the relaxation time, a quantity related to the entropy
production, making the process dissipative. In addition, since $\bM$ is even
under parity (P) while $\bE$ is odd, $\alpha_{ij}$ is zero for
parity invariant systems. The kME effect also vanishes in the
absence of spin-orbit interaction. In 2D n-doped inversion layers
with Rashba spin-orbit interaction, this effect has been predicted
more than a decade ago \cite{edelstein,aronov}, but has been
observed experimentally only very
recently \cite{kato, ganichev}.

In the model used by Levitov {\it et. al.}\cite{levitov},
the kME effect originates from an electron scattering by impurities
whose potential lacks inversion symmetry.
As such, the effect is extrinsic and actually {\it vanishes} in the clean
limit.

In contrast, we present an analysis of hole-doped
semiconductor without inversion symmetry where the spin-orbit
splitting of the p-band is intrinsic. In the absence of strain the system
is both T and P invariant and hence no kME effect occurs.
As we argue below, the shear strain induces a P-breaking term
in the Hamiltonian which is responsible for the effect.

There are several advantages to having a piezo-magneto-electric
effect in 3D p-doped semiconductors (such as GaAs, GaSb, InSb,
InGaAs, AlGaAS). Technologically, engineering of different original
strain architectures is a common procedure in today's semiconductor
applications. By taking place in a 3D bulk sample, rather than in a
2D sample, this effect allows (with specific strain configurations)
full spatial manipulation of the magnetic moment. Most importantly,
the effect occurs in p-doped semiconductors, and thus it allows
manipulation of the direction of the collective ferromagnetic moment
which develops in (p-doped) dilute magnetic semiconductors.

\begin{figure}
  \includegraphics[width=8.0cm]{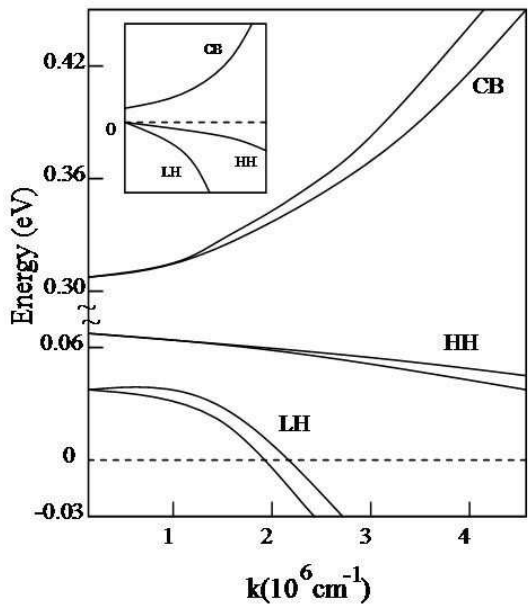}\\
  \caption{Dispersion curves for InSb at $4$ kbar stress on the $[110]$
  direction ($\epsilon_{xy} \approx 2 \times 10^{-3}$) and
  $\vec{k} \parallel [1 {\overline 1} 0]$
  as measured by Seiler \emph{et al} \cite{seiler}.
  The strain splits the conduction and the valence bands.
  The Dresselhaus $k^3$ type inversion asymmetry is
  negligibly small on this scale.
  No splitting is observed for strain in the $[001]$ direction.
} \label{bandstructure}
\end{figure}

Within the spherical approximation \cite{zhang03,lut}, the effective
Hamiltonian of a hole-doped semiconductor with spin-orbit coupling
is described by the Luttinger-Kohn model in the spin-$3/2$ band:
\begin{equation}\label{lut}
H_{LK}= \frac{1}{2m}\left(\gamma_1+\frac{5}{2}\gamma_2\right) \bk^2
- \frac{1}{m} \gamma_2 (\bk \cdot \bS)^2,
\end{equation}
where  $S_i$ is the spin-$3/2$ (4$\times$4 matrix) operator,
$\gamma_1$ and $\gamma_2$ are material-dependent
Luttinger constants. The band structure consists of a doubly
degenerate heavy hole band corresponding to $\hat{\bk}\cdot\bS = \pm 3/2$ and a
doubly degenerate light hole band with $\hat{\bk}\cdot\bS = \pm 1/2$
(see inset of Fig.(\ref{bandstructure})). The above
Hamiltonian is both $P$ and $T$ invariant.
The strain,  being a second order symmetric tensor $\eps_{ij}$,
naturally couples to $S_i S_j$ and to zero-th order modifies
the original Hamiltonian $H_{LK}$ by the term
\begin{equation}\label{lut}
H_{\epsilon}= D_d(\eps_{xx} + \eps_{yy}+ \eps_{zz})  + D_u
\epsilon_{ij} S_i S_j, \;\;\; i,j=x,y,z
\end{equation}
\noindent where $D_d$ and $D_u$ are the usual hydrostatic and shear
deformation potentials \cite{pikus}. The modified Hamiltonian
$H_{LK}+H_{\epsilon}$ remains invariant under both P and T. Each of
the two valence bands is still doubly degenerate. As seen in Fig.
(\ref{bandstructure}), the strained Hamiltonian exhibits a finite
energy gap between the heavy and light hole bands at zero momentum
\textbf{k}=0. External electric field will cause a spin current
\cite{zhang03}, but no uniform magnetization. For semiconductors
with inversion symmetry these are the only terms allowed at
quadratic order in $\bk$.

However, in the absence of an inversion symmetry center, the shear strain
induces a P-breaking term {\em linear} in momentum \cite{pikus}:
\begin{equation} \label{perturbhamilt}
H'=\lambda_i S_i, \;\;\; \lambda_x = C_4 (\epsilon_{xy} k_y -
\epsilon_{xz} k_z),
\end{equation}
\noindent where $\lambda_y, \lambda_z$ are obtained from $\lambda_x$
by cyclic permutation of indices and $C_4$ is a material constant
related to the interband-deformation potential for acoustic phonons.
This term is responsible for the piezo-kME effect. Its origin can be
traced back to the Kane's $8\times 8$ model ($2\times2$ for each the
conduction and the split-off band, and $4\times 4$ for the valence
band)(Fig.(\ref{bandstructure})), within which the valence band
couples to both the conduction band and the split-off band. Upon
straining, the zeroth order effect is the P-invariant coupling
mentioned in the previous paragraph. At the first order, the
conduction band $|s \rangle$ and the valence bands $|x\rangle, |y
\rangle, |z \rangle$ couple, and the matrix elements between the
valence and conduction band have the form $\epsilon_{xy} \langle s
|\partial_x
\partial_y | z \rangle$ (plus cyclic permutations) where $|s
\rangle$ is the $s$-orbital and $|z \rangle$ is one of the $p$
orbitals. Any other combination will not satisfy the $L_z$ selection
rule. In systems with inversion symmetry where the selection rules
for \textbf{L} are satisfied, it is impossible to couple the
spin-$0$ ($|s \rangle$) conduction states with spin-$1$ ($|x\rangle,
|y \rangle,|z \rangle$) valence states through a spin-$2$ term (rank
2 tensor) ($\epsilon_{ij}$) and hence $\langle s | \partial_x
\partial_y | z \rangle =0$.
However, when inversion symmetry is broken, $\langle s | \partial_x
\partial_y | z \rangle  \ne 0$ as the \textbf{L}
selection rule does not apply. We then obtain an $8\times 8$ Kane
matrix with the strain terms describing the interaction between
valence and conduction bands. To find an effective $4\times 4$
hamiltonian for the valence band, one must project onto the
valence band while taking into account the interactions with the
conduction and the split-off band. The first term which appears in
perturbation theory is the Hamiltonian (\ref{perturbhamilt}).
Reciprocally, a similar term will appear in the conduction band,
with the spin there being a spin-1/2 matrix. These terms have been
observed experimentally \cite{seiler} although recent evidence
suggests other effects could also play a role \cite{kato}.

The piezo-kME effect can be easily understood as follows:
assume a material strained only along the $[110]$ direction,
such that $\epsilon_{xy}\ne 0$ is the only nonvanishing shear
strain component. Hence, the P-breaking term in the hamiltonian is
$H' = C_4 (\epsilon_{xy}k_y S_x - \epsilon_{yx} k_x S_y)$.
This effectively corresponds to Zeeman
coupling of hole-spins with a fictitious internal magnetic field
$B_x = C_4\epsilon_{xy} k_y / \mu_B, \;\; B_y = -C_4 \epsilon_{xy} k_x /\mu_B$, $\mu_B$
being the Bohr magneton. Upon the application of an
electric field along, say, the $y$ axis, the average momenta become
$<k_x> \approx 0, \;\; k_y \approx eE\tau/m$ where $\tau$ is the momentum
relaxation time. In turn, this gives
$<B_y> \approx 0, \;\; <B_x> \approx C_4 \epsilon_{xy}eE\tau/m \mu_B$.
The non-zero $<B_x>$ field now couples to the spins and orients them along the $x$
axis. This gives rise to a magnetization perpendicular to the
electric field. Alternatively, the electric
field along the $x$ axis will induce magnetization along the $y$ axis of equal
modulus but of opposite sign to the previous one.
This has recently been observed in the conduction band by Kato {\it et.al.}
Ref.\cite{kato}.
Moreover, if we assume linear dependence on the relaxation time, $\tau$, and
neglect the effect of parity conserving strain term, then the
the form of $\alpha_{ij}$ is constrained by dimensional analysis
alone
\begin{equation}
  \alpha_{ij}=\mu_{B} n^{\frac{2}{3}} \frac{e\tau}{\hbar}\times\Phi_{ij}\!\!
  \left(\!\frac{m C_4}{\gamma_1 \hbar^2 n^{\frac{1}{3}}},\frac{\gamma_1}{\gamma_2
}\!  \right),
\end{equation}
where $n$ is the carrier density and  the scaling function
$\Phi_{ij}(x,y)$ vanishes linearly with its first argument $x$.
Based on the above argument, up to a sign, its components should be
proportional to $\eps_{ij}$.

\begin{figure}
 \begin{tabular}{c}
   \includegraphics[width=5.5cm]{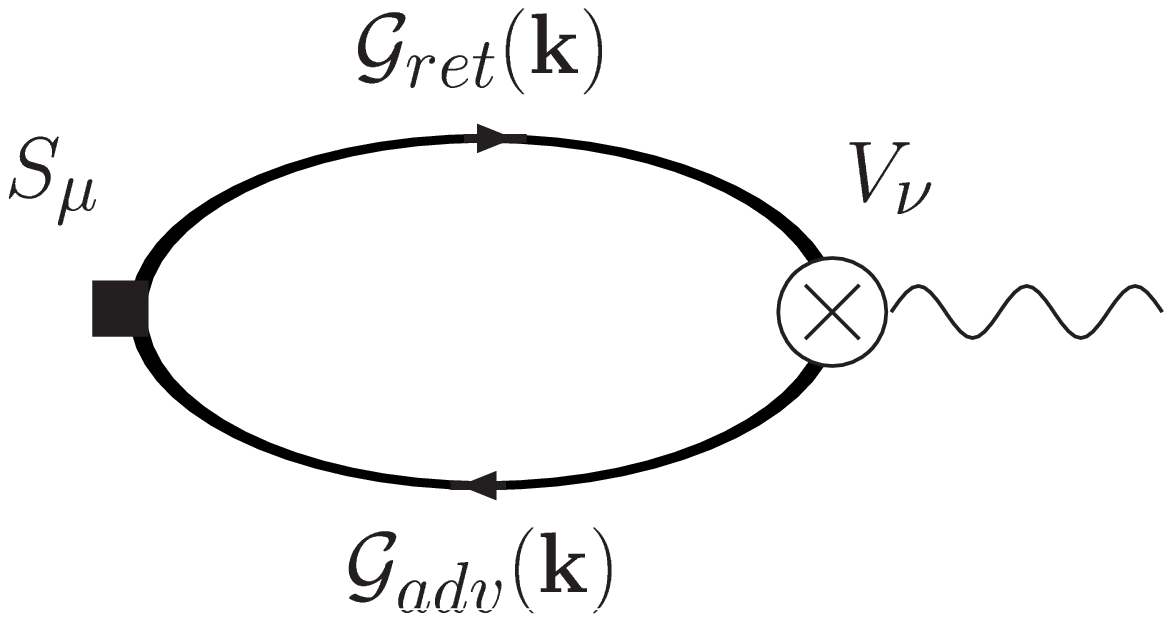}\\
   \includegraphics[width=8cm]{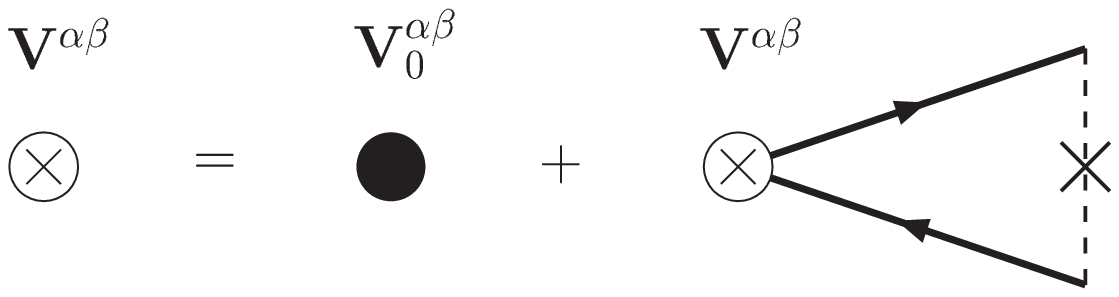}
 \end{tabular}
  \caption{Top: Feynman diagram representing the kinematic Magneto-Electric
effect in 3D. Bottom:
Kinetic equation for the vertex matrix in the ladder approximation. Here
${\bf V}_0^{\alpha\beta}$ is the velocity operator in the absence of impurities.
   }\label{bv}
\end{figure}

We shall now justify the above claims.
The static spin response to the d.c. electric field can be shown to be
given by
\begin{equation}
 \alpha_{\mu\nu}=\frac{\mu_B}{\hbar}
\lim_{\omega\rightarrow 0}\Im m \left[\frac{Q^{ret}_{\mu\nu}(\omega)}{\omega} \right],
\end{equation}
where $\mu_B$ is the Bohr magneton and the retarded correlation function
$Q^{ret}_{\mu\nu}(\omega)=Q_{\mu\nu}(i\omega \rightarrow \omega+i\eta)$,
($\eta \rightarrow 0^+$), and
\begin{equation}\label{qresp}
Q_{\mu\nu}(i\omega)=\int_{0}^{\beta}d\tau e^{i\omega \tau}\langle T S_{\mu}(\tau)j_{\nu}(0)
 \rangle.
\end{equation}
For d.c. response only spatial averages of the spin and the current
operators need to be considered above.
The corresponding diagram is shown in Fig.(\ref{bv}).

Since the strain splitting is typically small compared to the spin-orbit
splitting at the Fermi surface (Fig.\ref{bandstructure}),
we can include its effects within (degenerate) perturbation theory.
Utilizing the powerful mapping between spin-$3/2$ SU(2) and SO(5)
representations pioneered by Murakami, Nagaosa and Zhang
\cite{zhang03}, the unperturbed thermal Green's functions
can be conveniently written as
\begin{equation}\label{green}
G_0(\bk,i\omega_n)=\frac{1}{2}\sum_{s=\pm 1} \frac{1+s\hat{d}_j(\bk)
\Gamma_j}{-i\omega_n+(1+s\Delta)\eps(\bk)},
\end{equation}
where $\Delta=2\gamma_2/\gamma_1$, $\eps(\bk)=\gamma_1\bk^2/2m$,
$j=1,\ldots,5$ and $\hat{d}(\bk)$ is a (spherical) unit vector in the 5-dimensional
space, equivalently $\hat{d}_m(\bk)=Y^m_{l=2}(\theta_{\bk},\phi_{\bk})$ where $Y$'s are
spherical harmonics and the angles are in $\bk-$space; $\Gamma_j$ are 5 Dirac gamma
matrices \cite{zhang03}.
Upon inclusion of spinless impurities with potential $u(\bk)$,
the full (impurity) Green's function
$G(\bk,i\omega_n)=G_0(\bk,i\omega_n+\Sigma(\bk,i\omega_n))$.
Within the Born approximation the self-energy is
$\Sigma(\bk,i\omega_n)=n_{imp}\int\!\!d\bq \;|u(\bk-\bq)|^2 G_0(\bq,i\omega_n)$;
$n_{imp}$ is the concentration of impurities.
Finally, to leading order in P breaking strain \cite{neglect},
\begin{equation}\label{gfull}
{\cal G}(\bk,i\omega_n)=\left[1-G(\bk,i\omega_n)H'(\bk)\right]G(\bk,i\omega_n).
\end{equation}
Subsequently, all of the calculations will be carried out using (\ref{gfull}).

As shown in Fig.(\ref{bv}), the finite frequency response function
(\ref{qresp}) is given by
\begin{equation}
Q_{\mu\nu}(i\Omega) = -\frac{e}{\beta}\sum_{\omega_n}\int d\bk Tr\left[ S_{\mu}
\Pi_{\nu}(\bk,i\omega,i\Omega) \right],
\end{equation}
where the trace is over the heavy/light hole spaces and
where, as shown in Fig.(\ref{bv}), the ($4\times4$ matrix) vertex function $\Pi_{\nu}$
satisfies the kinetic equation (within the ladder approximation)
\begin{widetext}
\begin{equation}\label{ladder}
\Pi_{\mu}(\bk,i\omega,i\Omega)={\cal G}(\bk,i\omega)
V_{\nu}(\bk){\cal G}(\bk,i\omega-i\Omega)+
n_{imp} {\cal G}(\bk,i\omega)
\int \!\!d\bq |u(\bk-\bq)|^2 \Pi_{\nu}(\bq,i\omega,i\Omega)
{\cal G}(\bk,i\omega-i\Omega);
\end{equation}
\end{widetext}
The velocity operator $V_{\mu}(\bk)=\partial H(\bk)/\partial k_{\mu}$.
Note that ${\cal G}$ does not commute with $\Pi_{\nu}$. As such we have 16
coupled integral equations to solve, one for each entry of the 4$\times$4
matrix. In the case of $\delta$-function impurities $u(\bk-\bq)=u_0$ is a constant
and the above integral equation is separable. Integrating both sides over $\bk$,
it is easy to see that in the absence of parity breaking strain,
the vertex correction vanishes \cite{murakami}. On the other hand,
for finite strain, the vertex correction does not vanish, and we still have to
solve a system of 16 coupled equations.

However, to leading order in the strain it can be seen that {\em all 16 equations decouple
in the basis of the Clifford algebra}! Expanding the vertex matrix
\begin{equation}
\Pi_{\mu}(\bk,i\omega,i\Omega)= \sum_{A=0}^{15}
\lambda_{\mu}^{A}(\bk,i\omega,i\omega-i\Omega)\Gamma_A,
\end{equation}
where the sum runs over all 16 elements \cite{sakurai} and $\lambda_{\mu}^{A}(\bk,i\omega,i\omega-i\Omega)$
is now an ordinary vector function.
Since $Tr[\Gamma_A\Gamma_i\Gamma_B\Gamma_i]$ is diagonal in $A$ and $B$
it is easy to see that
$$
\frac{1}{4}\!\!\int\!\!d\bk Tr[\Gamma_A{\cal G}(\bk,i\omega)\Gamma_B{\cal G}(\bk,i\omega-i\Omega)]=
M_A(i\omega,i\Omega) \delta_{AB}
$$
Therefore,
\begin{eqnarray}\label{vertexsol}
\Pi_{\mu}(\bk,i\omega,i\Omega) \!\!=\!\!{\cal G}(\bk,i\omega) \bigl(
V_{\mu}(\bk)\!+\!R_{\mu}(i\omega,i\Omega) \bigr) {\cal
G}(\bk,i\omega\!-\!i\Omega).
\end{eqnarray}
where
\begin{equation}\label{R}
R_{\mu}(i\omega,i\Omega)=
n_{imp}u_0^2 \left(
\sum_{A=0}^{15}\frac{ \Gamma_A V^A_{\mu}(i\omega,i\Omega)}{1-n_{imp}u_0^2 M_A(i\omega,i\Omega)}
\right)
\end{equation}
and
\begin{equation}
V^A_{\mu}(i\omega,i\Omega)=\frac{1}{4}\!\!\int\!\!d\bk
Tr\left[\Gamma_A {\cal G}(\bk,i\omega)V_{\mu}(\bk){\cal
G}(\bk,i\omega-i\Omega)\right],
\end{equation}
\begin{equation}
M_A(i\omega,i\Omega)=\frac{1}{4}\!\!\int\!\!d\bk Tr\left[\Gamma_A
{\cal G}(\bk,i\omega)\Gamma_A{\cal G}(\bk,i\omega-i\Omega)\right].
\end{equation}

With the known structure of the vertex matrix (\ref{vertexsol}), we
can compute the response to the $E_{\mu}$ field. Following the
standard technique \cite{mahan90} we can perform the Matsubara
summation, let $i\Omega\rightarrow \Omega+i\eta$, take the limit of
$\Omega\rightarrow 0$ and finally take the temperature $T\rightarrow
0$ to find
\begin{equation}
\alpha_{\mu\nu} = e\int\!\!d\bk Tr[S_\mu {\cal
G}^{ret}(\bk)(V_{\nu}(\bk)+R_{\nu}){\cal G}^{adv}(\bk)],
\end{equation}
where the vertex matrix $R_{\mu}$ is given by the discontinuity of
(\ref{R}), $R_{\mu}=R_{\mu}(i\eta,-i\eta)$.
Finally, ignoring the interband transitions,
(i.e. in multiple sums over $s$ in (\ref{green}) we keep
only the same $s$), we get
\begin{equation}
\alpha_{ij}=-\mu_B n^{\frac{1}{3}}\frac{\partial \lambda_i}{\partial k_j} \frac{e\tau
}{\hbar^3}\frac{15}{2} \frac{3^{\frac{1}{3}}}{\pi^{\frac{1}{3}}} \frac{m}{\gamma_1}
\left(\sum_{s=\pm
1}\frac{1}{(1+s\Delta)^{3/2}}\right)^{\frac{2}{3}}
\end{equation}
where $\tau$ is the momentum relaxation time, $\mu_B=0.58\times
10^{-8}eV/G$ is the the Bohr magenton, $n$ is the carrier
concentration, and $\Delta=2\gamma_2/\gamma_1$. Since $\lambda_x =
C_4 (\epsilon_{xy} k_y - \epsilon_{xz} k_z)$ ($\lambda_y, \lambda_z$
being obtained through cyclic permuations of $x,y,z$) are linear in
the components $k_{x,y,z}$, the factor $\partial \lambda_i/ \partial
k_j$ is a momentum independent, strain dependent tensor. For GaAs,
$C_4 = C_3/2\eta$ where $C_3 = 8 \times 10^5 m/s$ \cite{pikus} is a
measured constant related to the deformation potential for acoustic
phonons while $\eta =E_{SO}/(E_g + E_{SO}) = 0.183$, $E_g$ being the
gap energy while $E_{SO}$ is the spin-orbit coupling energy for
GaAs. For GaAs $\gamma_1 = 6.98, \gamma_2=2.06$, hence $\Delta=
0.59$. This gives a value of $C_4/\hbar = 2.2 \times 10^6 m/s$. For
$n=3 \times 10^{16}cm^{-3}$, $\eps_{xy}=1 \%$, in a generic sample
of mobility $\mu = 50 cm^2/V \cdot s$ the magnetization due (and
perpendicular) to an electric field $E$ is $<S> = 5 \times 10^{16} E
(V^{-1} m^{-2})$. Under an electric field of $E = 10^4 V/m$ the
magnetization becomes $<S> = 6 \times 10^{14} cm^{-3}$. This
corresponds to almost ${<S>}/{n} = 2\%$ spin orientation efficiency.
Since $<S> \sim k_F \sim n^{1/3}$ the spin orientation efficiency
will grow for small hole concentration $n$.

An interesting new application of the piezo-kME effect would be to
manipulate the {\it collective} magnetization of the dilute magnetic
semiconductors \cite{ohno}. It is believed, at least in the high
mobility metallic regime, that the ferromagnetism of Mn$^{++}$ ions
in GaMnAs is hole mediated. Within the $\bk \cdot \bp$ method
\cite{dietl,macdonald}, the coupling between the collective
magnetization and the electric field induced spin polarization is of
the order $J \approx 1eV$ \cite{dietl,macdonald}. Thus, even if the
total electric field induced magnetization represents 1 Bohr
magneton per 10$^3$ spins, the effective magnetic field, felt by the
Mn spins is $H_{eff}\approx 17$Tesla!

We wish to thank Profs. S.C. Zhang and  D. Goldhaber-Gordon for useful
discussions. B.A.B. acknowledges support from the SGF Program. This
work is supported by the NSF under grant numbers DMR-0342832 and the
US Department of Energy, Office of Basic Energy Sciences under
contract DE-AC03-76SF00515.


\begin{thebibliography}{10}
\bibitem{lut} The deviation from spherical approximation is less
than $10\%-20\%$ for most materials.
\bibitem{neglect} We neglect the parity preserving strain term, 
which does not alter the conclusions dramatically.
\end{thebibliography}
\end{document}